# Scale-invariant projection optimization in tomographic volumetric additive manufacturing

Seungpyo Woo[1], Sangyup Lee[2], Hayden K. Taylor[1]

[1]Department of Mechanical Engineering, University of California, Berkeley.

Berkeley, CA 94720, USA

[2]H. Milton Stewart School of Industrial and Systems Engineering, Georgia Institute of Technology.

Atlanta, GA 30332, USA

**Abstract**

Tomographic volumetric additive manufacturing (TVAM) requires projection patterns that achieve high in-part fidelity while suppressing unintended exposure outside the target. We present a scale-invariant projection optimization framework (SiPO) that decouples projection shape from absolute dose scaling. The method formulates projection design as a linear-fractional program based on normalized conformity and spillage metrics, which is converted into a linear program via the Charnes–Cooper transformation. Two practical deterministic cases are introduced for process control: minimizing dose spillage under strict material tolerances and maximizing target conformity under hard inhibition constraints. A matrix-free primal-dual hybrid gradient solver enables large-scale implementation. Numerical results demonstrate that the framework provides a clear trade-off between target fidelity and process separation and remains effective under 3D blur-aware forward models.

## 1. Introduction

Tomographic volumetric additive manufacturing (TVAM) introduced a distinct route to three-dimensional fabrication by forming an entire volume through tomographically delivered light rather than by assembling geometry layer by layer. Since the original computed axial lithography demonstrations established the core concept of rapid, support-free volumetric printing, subsequent advances have improved spatial resolution and process control and expanded TVAM into a broader manufacturing platform [1], [2]. This expansion is evident not only in the maturation of the printing process itself, but also in the widening range of target applications. TVAM has been extended to bioprinting of complex living tissue constructs, highlighting its promise for soft and cell-laden materials [3], to in-space manufacturing, where its layer-less operation is attractive because it does not rely on maintaining a flat liquid–gas interface and can reduce gravity-driven distortion during fabrication [4], and more recently to shape-memory functional foams, where thermally expandable resins enabled lightweight, reprogrammable structures with dimensions exceeding the original resin vial size [5]. Together, these developments indicate that TVAM is no longer confined to being a rapid-printing novelty, but is increasingly emerging as a versatile platform for application-specific volumetric manufacturing across biomedical, aerospace, and functional-material systems.

As TVAM has expanded across a wider range of resins and target applications, it has become increasingly clear that the relevant design quantity is not optical dose itself, but the resulting material response. In practice, fabrication quality is governed by how delivered energy is converted into thresholded curing, conversion, refractive-index change, or other response-derived properties through a material-dependent law [6], [7], [8]. This response-oriented viewpoint is particularly important because process parameters such as penetration depth, curing threshold, temporal sampling, and exposure time can substantially affect print quality and robustness even in simplified energetic models of TVAM [9].

These developments make the inverse problem in TVAM increasingly central. One must determine projection patterns that generate the intended in-part response while suppressing unintended curing in the surrounding liquid region. Existing studies have approached this challenge from several directions. High-fidelity TVAM emphasized principled optimization of illumination patterns for improved geometric reproduction [10]. Object-space model optimization (OSMO) showed that direct object-domain formulations can improve dose contrast and grayscale capability relative to simple filtered-backprojection-based approaches [11]. More recently, band-constrained projection optimization made out-of-bounds exposure an explicit design quantity, while process-aware penalty formulations such as TVAM AID and OSPW clarified how threshold selection, process window, and in-part dose range depend on the optimization objective itself [12], [13].

A parallel line of work has pushed TVAM optimization toward more physically informed forward models. Deconvolution-based TVAM showed that blur and transport effects from chemical diffusion and optical point spread function (PSF) can be compensated through experimentally informed correction of the target or projection patterns [14]. More recently, Dr.TVAM reformulated TVAM pattern computation as an inverse rendering problem capable of explicitly modeling scattering media, absorptive decay, and non-standard vial geometries, thereby substantially broadening the class of physical effects that can be incorporated into pattern optimization [15]. These studies clearly demonstrate the importance of physical modeling in TVAM, but they also highlight a practical trade-off: as the forward model becomes more expressive, computational cost and algorithmic complexity increase, and the resulting optimization may become less directly interpretable in process-design terms.

Conceptually, TVAM projection design is closely related to fluence-map optimization (FMO) in intensity-modulated radiation therapy (IMRT), where one likewise seeks to concentrate dose in a prescribed region while suppressing exposure elsewhere. In IMRT and the broader operations research literature, linear programming and linear-fractional programming have long provided rigorous tools for balancing conformity and spillage, often with global optimality guarantees when the problem is posed in a convex form [16], [17]. Yet this viewpoint has not been fully exploited in TVAM. Existing formulations often remain tied to an absolute dose scale, depend on manual threshold search, or require repeated parameter sweeps and heuristic weighting to identify process-feasible operating conditions [10], [12], [13].

In this work, we propose a scale-invariant projection optimization framework (SiPO) for both binary and grayscale targets in TVAM. The central idea is to determine the projection shape independently of absolute physical scale, so that target conformity and out-of-bounds suppression can be expressed through normalized metrics that remain interpretable across operating conditions. More generally, we treat TVAM projection design under a prescribed linear physical forward model, so that the optimization framework remains compatible with both idealized backprojection and calibrated process operators accounting for attenuation, blur, or effective point-spread effects. Within this framework, we introduce a band-region formulation that regulates where non-target exposure is penalized, enabling the optimization to shape the allocation of sacrificial spillage rather than merely suppressing it uniformly. This viewpoint also allows process feasibility to be interpreted directly: depending on the formulation, the optimization either identifies whether prescribed tolerances are attainable or computes the best target conformity achievable under a hard inhibition condition.

The contributions of this work are fourfold. First, we formulate a unified scale-invariant optimization framework that accommodates both binary and grayscale prescriptions within a common mathematical structure. Second, by separating projection shape from absolute scaling, we reduce reliance on manual threshold search and isolate the geometric component of the optimization from post-scaling with physical calibration. Third, we show that the band region is not merely a bookkeeping device for non-target voxels, but a genuine design variable that governs process separation, sacrificial spillage, and target conformity. Fourth, we demonstrate that this framework naturally extends to blur-aware physical settings through a general linear operator viewpoint, thereby connecting interpretable optimization design to deconvolution-relevant and effective-PSF-based process models. Taken together, these results position SiPO as a bridge between rigorous inverse planning and physics-aware TVAM process design.

## 2. Methods

### 2.1. Framework overview

The overall framework of SiPO is illustrated in Fig. 1. The procedure consists of four sequential stages: spatial configuration, response prescription, shape optimization, and physical scaling. First, the spatial domain is partitioned based on the target geometry to explicitly handle the intended structure and artifact-prone boundaries, shown in Fig. 1(a). To decouple material non-linearity from the projection optimization, the target material response $\mathbf{m}_T$ is mapped to an equivalent target dose field $\mathbf{f}_T$ using the inverse material response function $\mathcal{M}^{-1}$ in Fig. 1(b). Subsequently, the problem is formulated as a normalized linear program to find the optimal scale-invariant dose shape $\tilde{\mathbf{f}}^*$ that concurrently maximizes target conformity and minimizes dose spillage in Fig. 1(c). Finally, a scalar post-scaling parameter $\alpha$ is calibrated to map the normalized dose distribution back to the absolute physical domain, yielding the final material response in Fig. 1(d). The detailed mathematical formulations for each step are elaborated in the following subsections.

### 2.2. Domain definition

To formulate the proposed optimization problem, we separate the computational domain into two coupled spaces, the object space and the projection space. The object space describes the discretized target slice and its surrounding non-target regions, whereas the projection space represents the set of angle-dependent beamlets that determine the accumulated dose distribution. This dual-domain representation explicitly establishes the object–projection correspondence underlying the subsequent LP-based formulation. Although the present formulation is developed on a 2D slice-based computational model, it provides a foundation for future generalization to integrated 3D optimization with layer-to-layer coupling.

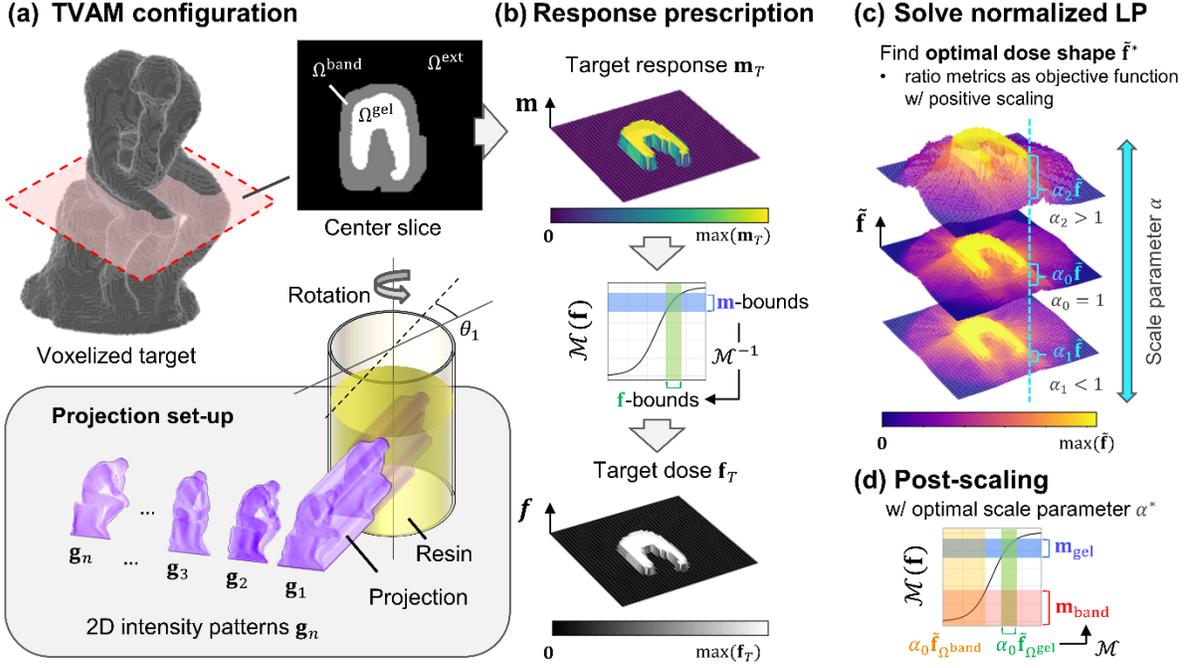

**Fig. 1.** Overview of the SiPO for TVAM concept. **(a)** TVAM configuration and spatial domain definition based on the voxelized target geometry. **(b)** Response prescription stage, converting the target material response $\mathbf{m}_T$ into the target dose $\mathbf{f}_T$ via the inverse material response function $\mathcal{M}^{-1}$ to preserve linearity. **(c)** Determination of the optimal scale-invariant dose shape $\tilde{\mathbf{f}}^*$ by solving the normalized linear program. **(d)** Post-scaling procedure determining the optimal scalar parameter $\alpha$ to recover the physical absolute dose and the corresponding material response.

The object domain shown in Fig. 2(a) is discretized on a uniform square grid of $N_{\text{obj},line} \times N_{\text{obj},line}$ elements, yielding the index set $\Omega = \{1, 2, \dots, N_{\text{obj}}\}$ with $N_{\text{obj}} = N_{\text{obj},line}^2$. Based on the prescribed target dose field $\mathbf{f}_T$, the object space is partitioned into three mutually exclusive regions. The gelation region $\Omega^{\text{gel}}$ consists of elements where the target dose is nonzero and material formation is intended. The band region $\Omega^{\text{band}}$ consists of non-target elements where the prescribed target dose is zero, but nonzero exposure may arise due to spatial spreading by the effective point spread function (PSF) operator $\mathbf{K}$. The exterior region $\Omega^{ext}$ consists of elements where neither target dose nor PSF-induced spill exists. This partition distinguishes the intended printed region from the surrounding artifact-prone boundary and the unaffected exterior background.

In parallel, the projection domain shown in Fig. 2(b) is discretized into $N_{\text{angle}}$ angular views and $N_{\text{beam}}$ beamlets per view, defining the projection index set $\Pi = \{1, 2, \dots, N_{\text{proj}}\}$ with $N_{\text{proj}} = N_{\text{beam}} N_{\text{angle}}$. Using the support of the forward projection $\mathbf{P}\mathbf{f}_T$, the projection space is further divided into an active set $\Pi^{\text{act}}$, containing rays that contribute to target formation, and a mask set $\Pi^{\text{mask}}$, containing rays that do not contribute to the gelation region and are therefore excluded from optimization. This object–projection domain configuration provides the discretized correspondence used in the subsequent LP formulation and enables the optimization to focus on physically relevant rays while explicitly monitoring out-of-bounds exposure in the band region.

Here, $i \in \Omega$ denotes an object-space element index and $j \in \Pi$ denotes a projection-space beamlet index. The prescribed target dose is denoted by $\mathbf{f}_T$, while $\Omega^{\text{gel}}$, $\Omega^{\text{band}}$, and $\Omega^{\text{ext}}$ denote the target, spill-prone boundary, and exterior background regions, respectively. Likewise, $\Pi^{\text{act}}$ and $\Pi^{\text{mask}}$ denote the active and masked beamlet sets in projection space. These notations are used throughout the following sections to define the forward operator and the optimization variables in a consistent object–projection framework.

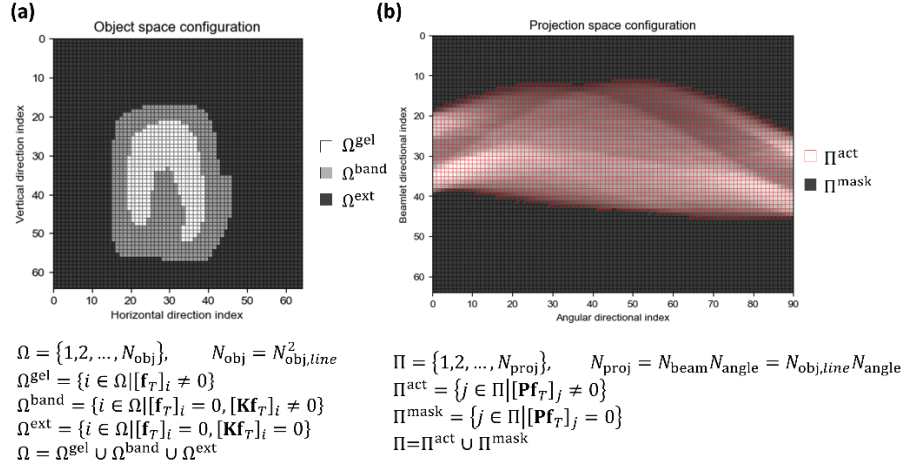

$\Omega = \{1,2,\dots,N_{obj}\}, \quad N_{obj} = N_{obj,line}^2$
$\Omega^{gel} = \{i \in \Omega | [\mathbf{f}_T]_i \neq 0\}$
$\Omega^{band} = \{i \in \Omega | [\mathbf{f}_T]_i = 0, [\mathbf{Kf}_T]_i \neq 0\}$
$\Omega^{ext} = \{i \in \Omega | [\mathbf{f}_T]_i = 0, [\mathbf{Kf}_T]_i = 0\}$
$\Omega = \Omega^{gel} \cup \Omega^{band} \cup \Omega^{ext}$

$\Pi = \{1,2,\dots,N_{proj}\}, \quad N_{proj} = N_{beam}N_{angle} = N_{obj,line}N_{angle}$
$\Pi^{act} = \{j \in \Pi | [\mathbf{Pf}_T]_j \neq 0\}$
$\Pi^{mask} = \{j \in \Pi | [\mathbf{Pf}_T]_j = 0\}$
$\Pi = \Pi^{act} \cup \Pi^{mask}$

**Fig. 2.** Domain definition for the SiPO framework. **(a)** Object space configuration partitioned into three mutually exclusive regions based on the prescribed target dose field: the gelation region $\Omega^{gel}$ intended for material formation, the band region $\Omega^{band}$ susceptible to spatial spreading caused by the effective PSF, and the unaffected exterior region $\Omega^{ext}$. **(b)** Projection space configuration divided into an active set $\Pi^{act}$ containing rays that contribute to target formation, and a mask set $\Pi^{mask}$ containing non-contributing rays that are strictly excluded from the optimization.

### 2.3. Forward operator

Following the general operator-based propagation notation adopted in the band constraint $Lp$-norm (BCLP) framework [12], the forward model in our study is first described in continuous form. Let $r$ denote the spatial coordinate in the object domain and $r'$ denote the coordinate in the projection domain. If $g(r')$ denotes the continuous projection function and $f(r)$ denotes the resulting accumulated dose field, the model is written as

$$f(r) = \mathcal{K}\mathcal{P}^*[g](r) \tag{1}$$

where $\mathcal{P}^*$ is the backprojection operator that maps the projection function to the object domain, and $\mathcal{K}$ denotes the effective object-space PSF operator. While the BCLP framework expresses the propagation process through a general operator-based description, the present formulation explicitly separates the additional object-space spreading effect through $\mathcal{K}$. This representation preserves linearity with respect to the projection function while allowing the forward model to account for effective spatial spreading in the object domain.

To apply the proposed optimization framework in the present study, the continuous forward model in Eq. (1) is expressed in the following discrete algebraic form.

$$\mathbf{f} = \mathbf{A}^T \mathbf{g} \tag{2}$$
$$\mathbf{A}^T = \mathbf{K}\mathbf{P}^T \tag{3}$$

where $\mathbf{g}$ is the projection vector (i.e., sinogram vector), $\mathbf{f}$ is the accumulated dose vector, $\mathbf{P}^T$ is the discrete backprojection (i.e., adjoint) operator, and $\mathbf{K}$ is the discrete effective PSF operator in object space. The composite operator $\mathbf{A}^T$ therefore represents the full dose deposition model from the projection domain to the object domain. Under the present slice-based formulation, this discrete operator provides the direct linear mapping used in the subsequent LP optimization.

Correspondingly, $\mathbf{P}$ denotes the forward projection operator, which is adjoint to $\mathbf{P}^T$ in the discrete setting. In this sense, $\mathbf{P}$ and $\mathbf{P}^T$ establish the object–projection operator pair, while $\mathbf{K}$ incorporates the effective spreading behavior after geometric backprojection. This continuous-to-discrete description is introduced to provide a physically interpretable representation of the forward model and to connect it to SiPO optimization used in the present study.

### 2.4. Response-to-dose mapping

In TVAM, the ultimate design objective is not merely to reproduce a prescribed dose distribution, but to realize a spatial distribution of material response, such as polymerization conversion, refractive index, or mechanical stiffness.

Let $\mathbf{m}_T$ denote the prescribed target response field. Since the relationship between the accumulated optical dose and the resulting material response is generally nonlinear, directly adopting the material response as the optimization target leads to the composite mapping $\mathbf{m} = \mathcal{M}(\mathbf{A}^T\mathbf{g})$, where $\mathcal{M}$ denotes the material response function. Such a formulation destroys the linear structure of the projection optimization problem and generally makes it difficult to preserve the tractability and convexity required for a linear-programming-based framework.

To decouple the material nonlinearity from the projection optimization, we first transform the prescribed target response $\mathbf{m}_T$ into an equivalent target dose field $\mathbf{f}_T$ using the inverse material response function $\mathcal{M}^{-1}$. That is, the target dose is defined as $\mathbf{f}_T = \mathcal{M}^{-1}(\mathbf{m}_T)$. With this definition, optimization no longer acts directly on the nonlinear response field but instead operates on the accumulated dose field $\mathbf{f}$ determined by the linear forward model $\mathbf{f} = \mathbf{A}^T\mathbf{g}$. In other words, the nonlinear material behavior is incorporated only in the pre-processing and post-processing stages, while the optimization core itself retains the linear operator structure introduced in Sec. 2.3.

In this work, the material response function $\mathcal{M}$ is modeled using a generalized logistic function, such as Richards' curve [18].

$$\mathbf{m} = \mathcal{M}(\mathbf{f}) = \alpha + \frac{k - \alpha}{\left[1 + e^{-\beta(\mathbf{f}-f_0)}\right]^{1/\gamma}} \tag{4}$$

where $\alpha$ and $k$ denote the lower and upper asymptotes, $\beta$ controls the response growth rate, $\gamma$ is a shape parameter, and $f_0$ is a location parameter characterizing the resin-specific response. This functional form is sufficiently flexible to represent monotonic photopolymerization behavior and, at the same time, provides an analytical inverse over the practical response range.

Accordingly, the inverse mapping from the target response to the target dose can be written as:

$$\mathbf{f}_T = \mathcal{M}^{-1}(\mathbf{m}_T) = f_0 - \frac{1}{\beta}\ln\left[\left(\frac{k - \alpha}{\mathbf{m}_T - \alpha}\right)^\gamma - 1\right] \tag{5}$$

This inverse transformation serves as a key link in the proposed SiPO framework. By converting a grayscale or property-based target into a dose-domain prescription prior to optimization, the subsequent optimization problem can be maintained as a linear problem with respect to the decision variable $\mathbf{g}$. After optimization, the computed dose field $\mathbf{f}$ can be mapped back through $\mathcal{M}$, allowing the solution to be interpreted in the physically meaningful material-response domain.

This formulation confines the physical nonlinearity to the pre-processing and post-processing stages. Consequently, the core optimization engine only evaluates the purely linear projection model $\mathbf{f} = \mathbf{A}^T\mathbf{g}$, thereby preserving the linearity and convexity required for LP-based formulation. Furthermore, as long as the target response lies within the invertible range of the selected material model, both binary and grayscale targets can be treated within a unified mathematical framework. Therefore, the proposed response-to-dose mapping serves as the key interface connecting nonlinear resin physics to the scale-invariant linear optimization strategy developed in the following sections.

## 2.5. Scale-Invariant Optimization Strategy and Metrics

In previous studies, performance metrics such as process window (PW) and in-part dose range (IPDR) have been used to evaluate reconstruction quality [8], [11], [13]. These metrics are useful for qualitatively and quantitatively assessing in-part dose uniformity or out-of-part artifact suppression in binary-target-based TVAM. However, when dealing with high-resolution grayscale targets or spatially varying target dose distributions, the existing metrics are insufficient to consistently evaluate both target conformity and out-of-bounds exposure. These conventional metrics were primarily designed for assessing the accuracy and uniformity of binary printing and do not fully reflect the additional evaluation requirements associated with real-valued response targets. Moreover, although a metric such as IPDR is useful for summarizing dose variation within a specific region, it does not directly connect relative conformity to the prescribed target distribution and worst-case dose spillage in the band region within a unified scale-invariant framework.

Accordingly, in the present work, we define a new metric based on a scale-invariant ratio form in order to avoid the numerical instability that arises when absolute dose bounds, $\mathbf{f}_L \leq \mathbf{f} \leq \mathbf{f}_U$, are directly imposed under the null space and ill-conditioning of the projection operator $\mathbf{A}^T = \mathbf{K}\mathbf{P}^T$, while simultaneously quantifying relative target conformity and band-region artifacts for grayscale targets. In this formulation, the optimization first determines the shape of the projection, whereas the absolute physical dose scale is selected later in the post-mapping stage.

In order to unify the general and application-specific formulations, we introduce $f_{\text{crit}}$ as an application-dependent critical reference dose, determined by the goal of each TVAM problem in practice. This scalar serves as the normalization anchor representing the most relevant dose scale for the manufacturing objective under consideration. Depending on the problem, $f_{\text{crit}} = \mathcal{M}^{-1}(m_{\text{crit}})$ may correspond to a representative target response $m_{\text{crit}}$, the minimum admissible lower bound for contrast-preserving fabrication, or a prescribed inhibition threshold in the non-target region.

To preserve the spatial target pattern up to a global positive scaling of the target dose field $\mathbf{f}_T$, the dose in the gelation region is normalized voxel-wise by the corresponding target dose. Based on this normalization, we define the dose-to-target value ratio (DTVR) as a measure of the spatial uniformity of the accumulated dose:

$$\text{DTVR} = \frac{\max\limits_{i \in \Omega^{\text{gel}}} \left( \frac{[\mathbf{A}^\mathrm{T}\mathbf{g}]_i}{\mathbf{f}_{T,i}} \right)}{\min\limits_{i \in \Omega^{\text{gel}}} \left( \frac{[\mathbf{A}^\mathrm{T}\mathbf{g}]_i}{\mathbf{f}_{T,i}} \right)} \tag{6}$$

Simultaneously, we define the Dose Spillage Ratio (DSR) to evaluate the worst-case out-of-bounds exposure in the band region $\Omega^{\text{band}}$. Since the dose in band region should be smaller than predefined critical dose $f_{\text{crit}}$, the numerator utilizes the absolute dose normalized by critical dose scalar $f_{\text{crit}}$, while the denominator shares the minimum relative dose from the gelation region:

$$\text{DSR} = \frac{\max\limits_{i \in \Omega^{\text{band}}} \left( \frac{[\mathbf{A}^\mathrm{T}\mathbf{g}]_i}{f_{\text{crit}}} \right)}{\min\limits_{i \in \Omega^{\text{gel}}} \left( \frac{[\mathbf{A}^\mathrm{T}\mathbf{g}]_i}{\mathbf{f}_{T,i}} \right)} \tag{7}$$

Both metrics exhibit zeroth-order homogeneity with respect to the projection vector $\mathbf{g}$. That is, scaling $\mathbf{g}$ by any positive constant $k > 0$ leaves the values of DTVR and DSR unchanged. Critically, the fact that both metrics share the same denominator, $\min\limits_{i \in \Omega^{\text{gel}}} \left( \frac{[\mathbf{A}^\mathrm{T}\mathbf{g}]_i}{\mathbf{f}_{T,i}} \right)$, is the cornerstone of the present formulation. This shared denominator simultaneously removes the global dose scale from both metrics, enabling them to be aggregated without introducing an additional scale-selection variable. Mathematically, this makes it possible to combine these two distinct engineering objectives into a single linear-fractional program (LFP), thereby preventing the computational complexity from escalating into a nonlinear multi-objective problem.

In this work, we adopt $f_{\text{crit}} = \min\limits_{i \in \Omega^{\text{gel}}} \mathbf{f}_{T,i}$ as a conservative reference scale for normalizing exposure in the band region. The target dose vector $\mathbf{f}_T$ is then normalized by the critical gelation dose $f_{\text{crit}}$ as $\tilde{\mathbf{f}}_T = \frac{\mathbf{f}_T}{f_{\text{crit}}}$, which gives $\tilde{f}_{\text{crit}} = 1$. Hence, DTVR and DSR can be rewritten as follows:

$$\text{DTVR} = \frac{\max\limits_{i \in \Omega^{\text{gel}}} \left( \frac{[\mathbf{A}^\mathrm{T}\mathbf{g}]_i}{\tilde{\mathbf{f}}_{T,i}} \right)}{\min\limits_{i \in \Omega^{\text{gel}}} \left( \frac{[\mathbf{A}^\mathrm{T}\mathbf{g}]_i}{\tilde{\mathbf{f}}_{T,i}} \right)}, \quad \text{DSR} = \frac{\max\limits_{i \in \Omega^{\text{band}}} [\mathbf{A}^\mathrm{T}\mathbf{g}]_i}{\min\limits_{i \in \Omega^{\text{gel}}} \left( \frac{[\mathbf{A}^\mathrm{T}\mathbf{g}]_i}{\tilde{\mathbf{f}}_{T,i}} \right)} \tag{8}$$

Although the numerator in Eq. (8) appears as an absolute dose, the denominator has dose units because $\tilde{\mathbf{f}}_T$ is dimensionless. Therefore, the rewritten DSR remains dimensionless and is algebraically equivalent to Eq. (7).

For experimental evaluation, additional metrics are reported to improve interpretability of band region exposure independently from gelation region.

### 2.6. Linear Programming via Two-Stage Relaxation

To simultaneously optimize out-of-bounds exposure and spatial uniformity while preserving the scale-invariant property, we formulate the objective function as a weighted sum of the numerators of DSR and DTVR over their shared denominator.

$$\min_{\mathbf{g}} \frac{w_1 \max\limits_{i \in \Omega^{\text{band}}} [\mathbf{A}^\mathrm{T}\mathbf{g}]_i + w_2 \max\limits_{i \in \Omega^{\text{gel}}} \left( \frac{[\mathbf{A}^\mathrm{T}\mathbf{g}]_i}{\tilde{\mathbf{f}}_{T,i}} \right)}{\min\limits_{i \in \Omega^{\text{gel}}} \left( \frac{[\mathbf{A}^\mathrm{T}\mathbf{g}]_i}{\tilde{\mathbf{f}}_{T,i}} \right)} \tag{9}$$

subject to

$$-\mathbf{g}_j \leq 0, \qquad \forall j \in \Pi^{\text{act}} \quad (10)$$
$$\mathbf{g}_j = 0, \qquad \forall j \in \Pi^{\text{mask}} \quad (11)$$

where, $w_1$ and $w_2$ are user-defined weighting factors that control the trade-off between dose spillage suppression and target conformity. A larger value of $w_1$ places greater emphasis on suppressing exposure in the band region, whereas a larger value of $w_2$ assigns greater importance to the relative dose uniformity within the gelation region.

In the objective function of Eq. (9), the first numerator term represents the maximum accumulated dose in the band region, which serves to suppress the worst-case magnitude of out-of-bounds exposure. The second numerator term represents the maximum relative dose within the gelation region, normalized by the target dose, and controls the upper-side conformity to the prescribed target pattern. Meanwhile, the denominator corresponds to the minimum relative dose in the gelation region, which represents the lowest relative dose level achieved within the gelation region. Therefore, the proposed objective function is designed to suppress both dose spillage in the band region and relative overdosing within the gelation region, while at the same time ensuring a sufficiently high minimum relative dose throughout the gelation region. In other words, the numerator acts to reduce unnecessary exposure and upper-side deviation, whereas the denominator acts to increase the lower dose bound in the gelation region. As a result, the formulation is constructed to simultaneously improve spatial uniformity and target conformity.

As shown in Eq. (10), the projection variables in the active region are constrained to take only nonnegative real values. This reflects the physical fact that optical intensity or projected dose cannot be negative. In addition, as specified in Eq. (11), the projection components in the mask region are fixed to zero, thereby eliminating rays that do not contribute to the gelation region and minimizing unnecessary dose spillage. Accordingly, the projection vector $\mathbf{g}$ takes only nonnegative real values: only $\mathbf{g}_j$ in the active region, where the corresponding ray passes through at least one gelation voxel, is allowed to be positive, whereas $\mathbf{g}_j = 0$ in the mask region to minimize dose spillage.

The resulting formulation contains two distinct sources of nonlinearity: the max-min operators and the fractional structure of the objective function. For this reason, the problem is relaxed in two stages. In the first stage, the extrema operators are replaced by auxiliary variables to obtain a standard LFP. In the second stage, the resulting fractional objective is transformed into an equivalent linear program. The details of these two steps are described in Sec. 2.6.1 and 2.6.2, respectively.

### 2.6.1. Extrema Relaxation for LFP

To resolve the nonlinearity induced by the max and min operators, the first stage of relaxation is performed. For this purpose, three auxiliary scalar variables, $u$, $v$, and $s$, are introduced. Here, $u$ upper-bounds the maximum absolute dose in the band region, $v$ upper-bounds the maximum relative dose in the gelation region, and $s$ lower-bounds the minimum relative dose in the gelation region. By introducing these auxiliary scalar variables, the nonlinear extrema constraints are replaced with linear constraints: $[\mathbf{A}^T \mathbf{g}]_i \leq u$ in the band region, and $s \leq \frac{[\mathbf{A}^T \mathbf{g}]_i}{\tilde{\mathbf{f}}_{T,i}} \leq v$ in the gelation region. Through this reformulation, the extrema operations are decoupled, and the original problem is transformed into a standard LFP.

$$\min_{\mathbf{g}, u, v, s} \frac{w_1 u + w_2 v}{s} \quad (12)$$

subject to

$$[\mathbf{A}^T \mathbf{g}]_i - u \leq 0, \qquad \forall i \in \Omega^{\text{band}} \quad (13)$$
$$[\mathbf{A}^T \mathbf{g}]_i - v \tilde{\mathbf{f}}_{T,i} \leq 0, \qquad \forall i \in \Omega^{\text{gel}} \quad (14)$$
$$s \tilde{\mathbf{f}}_{T,i} - [\mathbf{A}^T \mathbf{g}]_i \leq 0, \qquad \forall i \in \Omega^{\text{gel}} \quad (15)$$
$$-\mathbf{g}_j \leq 0, \qquad \forall j \in \Pi^{\text{act}} \quad (16)$$
$$\mathbf{g}_j = 0, \qquad \forall j \in \Pi^{\text{mask}} \quad (17)$$

Here, the objective function directly preserves the structure of the scale-invariant metric defined in the previous section, as shown in Eq. (12). Specifically, $u$ represents the worst-case dose spillage in the band region, $v$ represents the maximum relative dose within the gelation region, and $s$ represents the minimum relative dose level achieved in the gelation region. Accordingly, the present formulation is designed to suppress excessive exposure in the band region and relative overdosing within the gelation region, while at the same time maintaining a sufficiently high minimum relative dose throughout the gelation region.

That the accumulated dose at every voxel in the band region does not exceed the auxiliary variable $u$ is stated in Eq. (13), while the upper bound of the target-dose-normalized relative dose at every voxel in the gelation region is limited by $v$ can be seen in Eq. (14). In contrast, that the lower bound of the same relative dose is limited by $s$ is expressed in Eq. (15). Accordingly, $v$ and $s$ denote the upper and lower bounds, respectively, of the relative dose within the gelation region. Finally, to enforce the physical validity of the projection vector $\mathbf{g}$, a nonnegativity constraint is imposed in the active region and the mask region is fixed at zero, as presented in Eq. (16) and (17), which are identical to Eq. (10) and (11).

### 2.6.2. Charnes-Cooper Transformation to LP

The formulation derived in Eq. (12) successfully connects the proposed scale-invariant metrics to the optimization objective. However, the resulting problem is fundamentally a LFP with a nonlinear ratio objective, $\min_{\mathbf{x}} \frac{N(\mathbf{x})}{D(\mathbf{x})}$, where $\mathbf{x} = (\mathbf{g}, u, v, s)$, the numerator $N(\mathbf{x}) = w_1 u + w_2 v$, and the denominator $D(\mathbf{x}) = s$. In the present formulation, the feasible set is restricted to $s > 0$, which corresponds to a strictly positive relative dose within the gelation region.

In general, global optimization of fractional programs is challenging because of the nonlinear ratio structure. In the present case, however, the problem admits an exact transformation due to its specific functional form. After the extrema relaxation, the numerator $N(\mathbf{x})$ is an affine function associated with the upper-bound variables for the maximum operators, whereas the denominator $D(\mathbf{x})$ is an affine and strictly positive function associated with the lower-bound variable for the minimum operator over the feasible set. This structural property places the problem within the standard class of LFP for which exact reformulation techniques are available. Among the representative approaches for such problems, Dinkelbach's algorithm [19] and the Charnes–Cooper transformation (CCT) [20] are the two methods most considered.

Dinkelbach's algorithm is a robust iterative parametric method that converts the fractional program into a sequence of linear problems of the form $\max_{\mathbf{x}}(D(\mathbf{x}) - q_k N(\mathbf{x}))$, where $q_k$ denotes the ratio parameter updated at iteration $k$. At each iteration, the parameter is updated as $q_k = \frac{N(\mathbf{x}_k)}{D(\mathbf{x}_k)}$, and the algorithm terminates when $\max_{\mathbf{x}}(D(\mathbf{x}) - q_k N(\mathbf{x})) \approx 0$.

Although Dinkelbach's algorithm is attractive because of its superlinear convergence, it requires repeatedly solving a large-scale linear program constrained by the dense and ill-conditioned tomographic projection operator $\mathbf{A}^T$. In high-resolution TVAM, where the numbers of voxels and beamlets can reach the order of hundreds of millions, this repeated inner-outer iteration becomes a substantial computational burden.

For the present large-scale TVAM setting, the CCT is more suitable because it converts the LFP into an equivalent linear program through a single change of variables. Specifically, the CCT introduces a positive scalar variable $t = \frac{1}{D(\mathbf{x})} = \frac{1}{s}$, thereby normalizing the denominator to unity. The original variables are then mapped into the $t$-scaled domain as follows:

$$\mathbf{y} = t\mathbf{g}, \qquad \tilde{u} = tu, \qquad \tilde{v} = tv \tag{18}$$

Multiplying the constraints from Eq. (13) to Eq. (17) by the scaling variable $t > 0$ and substituting the transformed variables converts both the objective function and the feasible set into strictly linear forms. In particular, the lower-bound constraint $t(s\tilde{\mathbf{f}}_{T,i} - [\mathbf{A}^T\mathbf{g}]_i) \leq 0$ simplifies to $\tilde{\mathbf{f}}_{T,i} - [\mathbf{A}^T\mathbf{y}]_i \leq 0$ because $ts = 1$. Through this normalization, the denominator is eliminated from the objective without altering the underlying scale-invariant structure of the problem. As a result, the original LFP is exactly reformulated as the following linear program:

$$\min_{\mathbf{y},\tilde{u},\tilde{v}} w_1 \tilde{u} + w_2 \tilde{v} \tag{19}$$

subject to

$$[\mathbf{A}^T\mathbf{y}]_i - \tilde{u} \leq 0, \qquad \forall i \in \Omega^{\text{band}} \tag{20}$$
$$[\mathbf{A}^T\mathbf{y}]_i - \tilde{v}\tilde{\mathbf{f}}_{T,i} \leq 0, \qquad \forall i \in \Omega^{\text{gel}} \tag{21}$$
$$\tilde{\mathbf{f}}_{T,i} - [\mathbf{A}^T\mathbf{y}]_i \leq 0, \qquad \forall i \in \Omega^{\text{gel}} \tag{22}$$
$$-\mathbf{y}_j \leq 0, \qquad \forall j \in \Pi^{\text{act}} \tag{23}$$
$$\mathbf{y}_j = 0, \qquad \forall j \in \Pi^{\text{mask}} \tag{24}$$

It should be emphasized that the resulting LP does not remove the scale invariance of the original formulation. Rather, it provides a normalized representation of the underlying scale-invariant LFP by fixing the scaling gauge through normalization of the denominator to unity. Consequently, the LP determines only the optimal projection shape, $\mathbf{y}^*$, and equivalently the induced normalized dose field, $\tilde{\mathbf{f}}^* = \mathbf{A}^T\mathbf{y}^*$, up to an arbitrary positive scalar factor.

### 2.6.3. Post-scaling for Physical Projection

To recover a physically meaningful solution, a post-scaling step is required. Specifically, the physical projection vector $\mathbf{g}^*$ and dose field $\mathbf{f}^*$ are obtained as:

$$\mathbf{g}^* = \alpha^* \mathbf{y}^*, \qquad \mathbf{f}^* = \alpha^* \tilde{\mathbf{f}}^* \tag{25}$$

where, the optimal scaling factor $\alpha^* > 0$ is determined by the following one-dimensional calibration problem:

$$\alpha^* = \underset{\alpha > 0}{\operatorname{argmin}} \; \Phi(\alpha; \tilde{\mathbf{f}}^*) \tag{26}$$

In this study, the global scaling factor is determined using a weighted least-squares calibration functional. When the calibration is performed in the dose domain, the objective is defined as:

$$\Phi_f(\alpha) = \sum_{i \in \Omega^{\text{gel}}} \mathbf{w}_i \left( \alpha \tilde{\mathbf{f}}_i^* - \mathbf{f}_{T,i} \right)^2 \tag{27}$$

where, $\mathbf{w}$ denotes a nonnegative weight vector that specifies the relative importance of individual voxels in the gelation region. Under this formulation, the optimal scaling factor admits a closed-form solution $\alpha^* = \frac{\langle \tilde{\mathbf{f}}^*, \mathbf{f}_T \rangle_{\mathbf{W}}}{\langle \tilde{\mathbf{f}}^*, \tilde{\mathbf{f}}^* \rangle_{\mathbf{W}}}$, where $\mathbf{W} = \operatorname{diag}(\mathbf{w})$. Alternatively, the scaling factor may be determined in the material-response domain by minimizing the discrepancy after applying the nonlinear material response mapping:

$$\Phi_m(\alpha) = \sum_{i \in \Omega^{\text{gel}}} \mathbf{w}_i \left( \mathcal{M}(\alpha \tilde{\mathbf{f}}_i^*) - \mathbf{m}_{T,i} \right)^2 \tag{28}$$

Because $\mathcal{M}$ is nonlinear, this formulation leads to a one-dimensional nonlinear optimization problem. Nevertheless, the problem remains computationally inexpensive, since only a single scalar variable $\alpha$ is optimized.

This post-scaling formulation explicitly separates the geometric optimization of the dose distribution from the determination of its physically meaningful absolute scale. The former is handled by the LP in the normalized domain, whereas the latter is determined through a scalar calibration procedure. Such a decoupling provides a clear separation between geometric feasibility and material physics: the LP enforces the spatial structure of the solution, while the post-scaling step incorporates application-specific physical objectives through the choice of calibration functional.

### 2.7. Practical Formulations for Process Control

The generalized LP-based formulation is theoretically useful because it enables spatial uniformity and artifact suppression to be addressed within a single integrated optimization framework. By adjusting the weighting coefficients $w_1$ and $w_2$, the relative importance of these two objectives can be continuously reflected, which provides the advantage of exploring various solution trends and the associated trade-off relationships. However, in practical manufacturing environments, such a generalized optimization structure may have limitations when used directly as a process design criterion. In industrial TVAM processes, depending on the application, a more direct design requirement may be whether the final output satisfies pre-specified absolute physical criteria or tolerances, rather than exploring relative trade-offs. For example, quantities such as variations in optical opacity, deviations in mechanical properties, or the maximum allowable exposure in unintended regions may be interpreted not as matters of relative performance comparison, but rather as process constraints that must be satisfied.

From this perspective, in practical process control, a more deterministic formulation that directly incorporates given physical tolerances or performance criteria as constraints may be more appropriate than an approach based on exploring solution trends by varying the weighting coefficients $w_1$ and $w_2$. Accordingly, in this section, the generalized optimization framework introduced above is reformulated into several specific cases that can respond more directly to practical manufacturing requirements. In doing so, we aim to clarify the connection between the generalized theoretical framework and its practical implementation, and to present the proposed method in a way that can be more readily applied to TVAM process design and control problems.

### 2.7.1. Case 1: Minimizing Dose Spillage under Strict Material Tolerance

In many practical TVAM applications, the primary engineering objective is to suppress out-of-bounds exposure as much as possible, provided that the resulting material properties remain strictly within prescribed quality bounds. Let $\epsilon_L$ and $\epsilon_U$ denote the allowable lower and upper relative tolerances in the material-response domain for each voxel in the gelation region. The corresponding lower and upper response bounds are then defined as $\mathbf{m}_L = (1 - \epsilon_L)\mathbf{m}_T$ and $\mathbf{m}_U = (1 + \epsilon_U)\mathbf{m}_T$, respectively.

To preserve the linearity of the optimization solver, these physical response tolerances must be mapped back to the absolute dose domain through the inverse material-response function $\mathcal{M}^{-1}$ introduced in Eq. (5). This mapping yields the absolute dose bounds $\mathbf{f}_L = \mathcal{M}^{-1}(\mathbf{m}_L)$ and $\mathbf{f}_U = \mathcal{M}^{-1}(\mathbf{m}_U)$:

$$\mathbf{f}_{L,i} \leq \mathbf{f}_i \leq \mathbf{f}_{U,i}, \qquad \forall i \in \Omega^{\text{gel}} \tag{29}$$

To incorporate these absolute dose constraints efficiently into the normalized $\mathbf{y}$-space formulation, the critical gelation dose $\mathbf{f}_{\text{crit}}$ is redefined not by the prescribed target response, but by the absolute minimum allowable response required to form the weakest part of the gelation region.

$$m_{crit} = \min_{i \in \Omega^{\text{gel}}} \mathbf{m}_{L,i}, \qquad f_{\text{crit}} = \min_{i \in \Omega^{\text{gel}}} \mathbf{f}_{L,i} \tag{30}$$

Using this updated reference scale, the lower and upper dose bounds can be normalized as follows:

$$\tilde{\mathbf{f}}_{L,i} = \frac{\mathbf{f}_{L,i}}{f_{\text{crit}}}, \qquad \tilde{\mathbf{f}}_{U,i} = \frac{\mathbf{f}_{U,i}}{f_{\text{crit}}}, \qquad \forall i \in \Omega^{\text{gel}} \tag{31}$$

Under this strict specification, target conformity is no longer treated as a soft objective balanced by weighting factors. Instead, the admissible material-response range is enforced explicitly through hard lower and upper dose constraints in the gelation region. Accordingly, the generalized LP formulation naturally reduces to a deterministic LP with $w_1 = 1$ and $w_2 = 0$, whose sole objective is to minimize the normalized maximum dose spillage in the band region, represented by $\tilde{u}$. The solver is therefore dedicated entirely to suppressing out-of-bounds exposure, thereby securing the widest possible process window against artifacts.

$$\min_{\mathbf{y}, \tilde{u}} \tilde{u} \tag{32}$$

subject to

$$[\mathbf{A}^T \mathbf{y}]_i - \tilde{u} \leq 0, \qquad \forall i \in \Omega^{\text{band}} \tag{33}$$
$$[\mathbf{A}^T \mathbf{y}]_i - \tilde{\mathbf{f}}_{U,i} \leq 0, \qquad \forall i \in \Omega^{\text{gel}} \tag{34}$$
$$\tilde{\mathbf{f}}_{L,i} - [\mathbf{A}^T \mathbf{y}]_i \leq 0, \qquad \forall i \in \Omega^{\text{gel}} \tag{35}$$
$$-\mathbf{y}_j \leq 0, \qquad \forall j \in \Pi^{\text{act}} \tag{36}$$
$$\mathbf{y}_j = 0, \qquad \forall j \in \Pi^{\text{mask}} \tag{37}$$

Here, the objective function is constructed to minimize the worst-case normalized dose spillage in the band region as in Eq. (32). That the accumulated dose at every voxel in the band region does not exceed the auxiliary variable $\tilde{u}$ is stated in Eq. (33). The upper and lower admissible dose bounds in the gelation region are imposed by Eq. (34) and (35), respectively, thereby ensuring that the optimized dose field remains within the prescribed tolerance window throughout the gelation region.

Unlike the generalized relative formulation, the present deterministic LP explicitly restricts the feasible set through hard material constraints. Therefore, if the allowable dose interval defined by $\tilde{\mathbf{f}}_L$ and $\tilde{\mathbf{f}}_U$ is too narrow to be realized under the ill-conditioned projection operator $\mathbf{A}^T$, the LP is infeasible. In this sense, infeasibility itself provides useful diagnostic information, indicating that the requested material specification cannot be satisfied under the current hardware configuration, resin response, or dose-spreading characteristics.

Note that, in contrast to the general scale-invariant formulation, the specific formulation in this section introduces an explicit physical anchor $f_{\text{crit}}$ through absolute dose constraints. As a result, the global scaling factor for this specific use case is directly prescribed as $\alpha = f_{\text{crit}}$, and the physical projection and dose fields are simplified from Eq. (25) as follows:

$$\mathbf{g}^* = f_{\text{crit}} \mathbf{y}^*, \qquad \mathbf{f}^* = f_{\text{crit}} \tilde{\mathbf{f}}^* \tag{38}$$

This eliminates the need for post-scaling.

### 2.7.2. Case 2: Maximizing Target Conformity under Hard Inhibition Constraints

Alternatively, certain TVAM applications such as the fabrication of highly transparent optical components or delicate microfluidic structures require strict suppression of any void-region polymerization as the absolute priority. In such cases, even a small amount of unintended gelation in the surrounding band region may be unacceptable. Accordingly, the engineering objective shifts from dose-spillage minimization under target-quality tolerances to maximizing grayscale target conformity, that is, minimizing DTVR, while strictly enforcing a hard inhibition constraint in the band region.

Let $f_{\text{crit}}$ denote a predefined scalar critical dose, such as an oxygen-inhibition threshold, below which the liquid resin remains uncured. To prevent polymerization in the band region, the accumulated absolute dose must satisfy $\mathbf{A}^T\mathbf{g} \leq f_{\text{crit}}$. To incorporate this hard scalar bound into the normalized optimization framework, $f_{\text{crit}}$ is adopted as the global normalization constant. The normalized target dose field is then defined as follows:

$$\tilde{\mathbf{f}}_{T,i} = \frac{\mathbf{f}_{T,i}}{f_{\text{crit}}}, \qquad \forall i \in \Omega^{\text{gel}} \tag{39}$$

Under the predetermined condition, the physical projection vector $\mathbf{g}^*$ is recovered from the normalized LP solution by a deterministic scalar multiplication in Eq. (38). Accordingly, the corresponding absolute band dose is bounded by $\mathbf{A}^T\mathbf{g}^* = f_{\text{crit}}\mathbf{A}^T\mathbf{y}^* \leq f_{\text{crit}}\tilde{u}$. Therefore, to enforce the hard inhibition condition $\mathbf{A}^T\mathbf{g} \leq f_{\text{crit}}$, it is sufficient to impose the normalized constraint $\tilde{u} \leq 1$. Under this condition, dose spillage is no longer treated as a soft objective to be minimized. Instead, the generalized LP formulation reduces naturally to a deterministic LP with $w_1 = 0$ and $w_2 = 1$, whose sole objective is to minimize the worst-case relative overshoot within the gelation region, represented by $\tilde{v}$.

$$\min_{\mathbf{y},\tilde{v}} \tilde{v} \tag{40}$$

subject to

$$[\mathbf{A}^T\mathbf{y}]_i - 1 \leq 0, \qquad \forall i \in \Omega^{\text{band}} \tag{41}$$
$$[\mathbf{A}^T\mathbf{y}]_i - \tilde{v}\tilde{\mathbf{f}}_{T,i} \leq 0, \qquad \forall i \in \Omega^{\text{gel}} \tag{42}$$
$$\tilde{\mathbf{f}}_{T,i} - [\mathbf{A}^T\mathbf{y}]_i \leq 0, \qquad \forall i \in \Omega^{\text{gel}} \tag{43}$$
$$-\mathbf{y}_j \leq 0, \qquad \forall j \in \Pi^{\text{act}} \tag{44}$$
$$\mathbf{y}_j = 0, \qquad \forall j \in \Pi^{\text{mask}} \tag{45}$$

Here, the objective function is constructed to minimize the maximum relative dose overshoot in the gelation region as given in Eq. (40). That the normalized accumulated dose in the band region does not exceed unity, thereby satisfying the hard inhibition condition, is stated in Eq. (41). The upper and lower relative dose bounds in the gelation region are imposed by Eq. (42) and (43), respectively, thereby ensuring that the optimized dose field follows the prescribed normalized target pattern while remaining bounded from below throughout the gelation region.

Similar to Case 1, the proposed deterministic LP guarantees that the physical requirement in the band region is enforced directly within the linear solver. The optimal solution $\tilde{v}^*$ therefore provides a direct quantitative measure of the worst-case relative overshoot required to reproduce the target pattern without violating the inhibition threshold. If the target geometry or the inherent dose-spreading behavior makes the hard inhibition condition impossible to satisfy, the LP becomes infeasible. In this sense, infeasibility again serves as a practically meaningful diagnostic indicator, revealing that the requested structure cannot be realized under the current hardware configuration, material response, or dose-diffusion characteristics.

Although the strict lower-bound condition in Eq. (43) directs the solver to minimize the maximum relative overshoot $v$, it inevitably introduces a systematic positive bias across the gelation region. However, this limitation is analytically resolved by leveraging the scale-invariant nature of the SiPO framework. By directly applying the post-scaling weighted least-squares calibration defined in Eq. (27) and (28), the physical dose is recovered and corrected. Because the normalized dose is inherently bounded below by the prescribed target as $\tilde{\mathbf{f}}^* \geq \tilde{\mathbf{f}}_T$, this calibration functional mathematically guarantees a down-scaling factor where $\alpha^* \leq \mathbf{f}_{\text{crit}}$. Consequently, this post-scaling operation effectively centers the dose distribution around the target $\mathbf{f}_T$, maximizing true grayscale target conformity while strictly preserving the hard inhibition constraint in the band region.

To visually summarize the fundamental differences between the generalized formulation and the two practical process control cases, Fig. 3 illustrates how the optimization objectives and constraints interact with the nonlinear material response curve. In the general form in Fig. 3(a), the optimization concurrently minimizes the upper bounds for both the out-of-bounds dose $u$ and the relative target dose overshoot $v$ without hard absolute boundaries, exploring the trade-off governed by weights $w_1$ and $w_2$. In contrast, Case 1 in Sec. 2.7.1 strictly binds the dose within gelation region $\Omega^{\text{gel}}$ to the absolute limits $f_{L,U}$ derived from the material tolerance $m_{L,U}$, dedicating the solver entirely to minimizing the dose spillage $u$ in band region $\Omega^{\text{band}}$ as shown in Fig. 3(b). Conversely, Case 2 in Sec. 2.7.2 enforces a hard critical threshold $f_{\text{crit}}$ on the maximum allowable dose in band region $\Omega^{\text{band}}$ to strictly inhibit unintended gelation, while the solver minimizes the target conformity error $v$ within gelation region $\Omega^{\text{gel}}$ in Fig. 3(c). This comparative illustration clarifies how the flexible scale-invariant framework can be deterministically constrained to meet specific manufacturing tolerances.

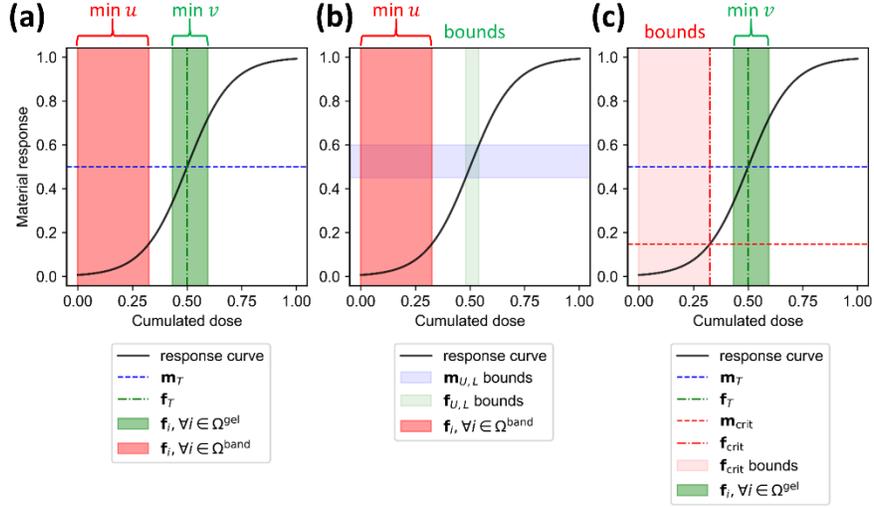

**Fig. 3.** Schematic representation of the optimization objectives and constraints mapped onto the material response curve for the three LP formulations. **(a)** General LP form: minimizing the dose spillage $u$ and relative overshoot $v$ as a weighted sum. **(b)** Case 1: minimizing dose spillage $u$ under strict material tolerance $f_{L,U}$, translated from the material response tolerances $m_{L,U}$ **(c)** Case 2: minimizing the relative overshoot $v$ under a hard absolute inhibition constraint $f_{\text{crit}}$ in the band region.

## 2.8. Computational Implementation via Matrix-Free PDHG

Although the CCT elegantly reformulates the scale-invariant metric into a standard linear program from Eq. (19) to (24), solving this LP using standard commercial solvers (e.g., based on interior-point methods or the simplex algorithm) is computationally prohibitive. High-resolution TVAM makes the explicit construction of the dense tomographic projection matrix $\mathbf{A}^T$ practically impossible due to memory constraints. To overcome this computational bottleneck, we adopted the primal-dual hybrid gradient (PDHG) algorithm, a first-order optimization method that operates entirely matrix-free, coded in PyTorch. The PDHG algorithm evaluates the forward operator $\mathbf{A}^T$ and adjoint operator $\mathbf{A}$ on-the-fly using operator-based computations, efficiently bypassing the memory limitations. The iterative primal descent and dual ascent steps of the implemented PDHG solver are directly governed by the Karush-Kuhn-Tucker (KKT) conditions of our formulated LP, whose detailed derivations and theoretical insights are provided in Supplementary Sec. 6.1.

## 3. Results and Discussion

To systematically evaluate the proposed scale-invariant linear programming (LP) framework, we consider a series of experiments spanning binary and grayscale targets, different band-region configurations, and both 2D and 3D forward models. The goal of this section is twofold: (i) to characterize the behavior of the proposed LP formulations under different design objectives, and (ii) to quantify the trade-off between in-part conformity and out-of-bounds dose separation.

As introduced in Sec. 2, the optimization framework is governed by two scale-invariant metrics: the dose-to-target value ratio (DTVR), which quantifies in-part dose uniformity, and the dose spillage ratio (DSR), which measures the relative severity of out-of-bounds exposure. While these metrics are well suited for optimization due to their shared denominator and compatibility with the linear-fractional formulation, DSR is less interpretable after post-scaling because it intrinsically couples band-region exposure with the minimum relative dose attained in the gelation region. To provide a more interpretable measure of process feasibility, we additionally introduce the process separation ratio (PSR), which directly compares the minimum value in the gelation region with the maximum value in the band region without relying on a case-dependent reference dose. Specifically, the PSR is defined as:

$$\text{PSR}_f = \frac{\min_{i \in \Omega^{\text{gel}}}[\mathbf{A}^T\mathbf{g}]_i}{\max_{i \in \Omega^{\text{band}}}[\mathbf{A}^T\mathbf{g}]_i}, \qquad \text{PSR}_m = \frac{\min_{i \in \Omega^{\text{gel}}}\mathcal{M}([\mathbf{A}^T\mathbf{g}]_i)}{\max_{i \in \Omega^{\text{band}}}\mathcal{M}([\mathbf{A}^T\mathbf{g}]_i)} \qquad (46)$$

where $\text{PSR}_f$ and $\text{PSR}_m$ denote the separation ratios evaluated in the dose and material-response domains, respectively. In contrast to DSR, PSR isolates the intrinsic separation between the gelation and band regions, thereby providing a direct, threshold-independent indicator of process feasibility.

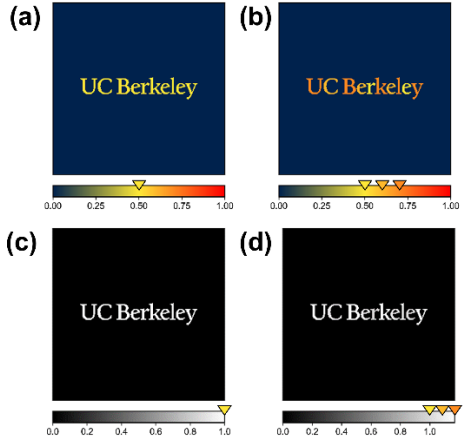

**Fig. 4.** Target configurations used in the experiments. **(a)**, **(c)** Binary UC Berkeley logo, and **(b)**, **(d)** grayscale UC Berkeley logo with letter-wise segmentation assigning three distinct target levels. The first row shows the target material response $m_T$, while the second row shows the corresponding normalized target dose $\tilde{f}_T$. The normalization is performed using a critical dose defined as $f_{\text{crit}} = \min_{i \in \Omega^{\text{gel}}} f_{T,i}$, such that all target doses are expressed relative to the minimum in-part dose. Target levels are indicated on the colorbars.

Using these complementary metrics, we evaluate the proposed formulations through a sequence of experiments of increasing complexity. Sec. 3.1–3.3 focus on two-dimensional optimization problems based on a $512 \times 512$ UC Berkeley logo as shown in Fig. 4, including both binary and grayscale target prescriptions, to characterize the baseline behavior of the LP formulations, their performance under spatially varying targets, and the effect of band-region definition. The grayscale UC Berkeley logo is constructed by assigning distinct target response levels to each segmented character in "UCBerkeley," thereby generating a spatially varying prescription within the gelation region given by (0.7, 0.6, 0.7, 0.6, 0.5, 0.7, 0.6, 0.7, 0.5, 0.7) as shown in Fig. 4(c).

In these experiments, the forward model is simplified by assuming an identity effective PSF operator, $\mathbf{K} = \mathbf{I}$, such that the composite operator reduces to $\mathbf{A}^T = \mathbf{P}^T$, allowing the intrinsic behavior of the optimization formulations to be isolated without additional diffusion or blur. The projection domain is discretized using 360 uniformly spaced angular views with a 1 degree interval. For the general formulation, the objective weights are fixed as $w_1 = 1.0$ and $w_2 = 1.0$, while the remaining experimental conditions are specified in each subsection.

Finally, Sec. 3.4 extends the analysis to a physically realistic three-dimensional setting using a Thinker target of size $128 \times 128 \times 130$ voxels, intended for actual printing scenarios. In this case, the forward model incorporates an effective diffusion-like PSF represented by a $21 \times 21 \times 21$ kernel, where only the central $5 \times 5 \times 5$ region is populated with a symmetric Gaussian profile with standard deviation $\sigma = 1.0$. This setup enables evaluation of the proposed framework under the full operator $\mathbf{A}^T = \mathbf{K}\mathbf{P}^T$, capturing the impact of spatial spreading on dose distribution and process feasibility.

### 3.1. Binary target: baseline behavior of LP formulations

For the binary benchmark in Fig. 5, the prescribed target response in the gelation region was fixed at $m_T = 0.5$. The three LP formulations were then compared under their respective design objectives: the general formulation, the target-bounded formulation (Case 1), and the dose-shaping formulation (Case 2). In Case 1, the admissible response interval was defined by a 10% tolerance: $m_L = 0.45$ and $m_U = 0.55$. In Case 2, the critical response threshold was set to $m_{\text{crit}} = 0.23$ after observing the maximum band responses obtained from the general formulation and Case 1, such that the inhibition condition lay between these two preceding outcomes. This choice was intended to establish an intermediate hard-inhibition condition for evaluating how much target conformity could be recovered under a stricter band constraint than the general formulation, but a less conservative one than the spillage-minimizing Case 1, as summarized in Fig. 5(a–e).

Among the three formulations, the general LP produced the highest target conformity. As shown in Fig. 5(c,d), the signed response deviation was tightly concentrated around zero, and the corresponding uniformity metrics reached nearly ideal values, with $\text{DTVR}_m = 1.002$ and $\text{DTVR}_f = 1.001$. This behavior indicates that, for a binary target, the weighted general formulation effectively recovers a target-centered solution after post-scaling. However, this superior conformity was accompanied by weaker gel–band separation than in the two deterministic cases. Specifically, the separation metrics were $\text{PSR}_m = 1.95$ and $\text{PSR}_f = 1.27$, while the maximum values in the band region remained relatively high at $m^*_{\text{band,max}} = \max_{i \in \Omega^{\text{band}}} \mathbf{m}^*_i = 0.26$ and $f^*_{\text{band,max}} = \max_{i \in \Omega^{\text{band}}} \mathbf{f}^*_i = 0.39$. Consistent with the histogram in Fig. 5(e), the general formulation therefore favors accurate in-part reconstruction while tolerating a larger level of out-of-bounds dose spillage.

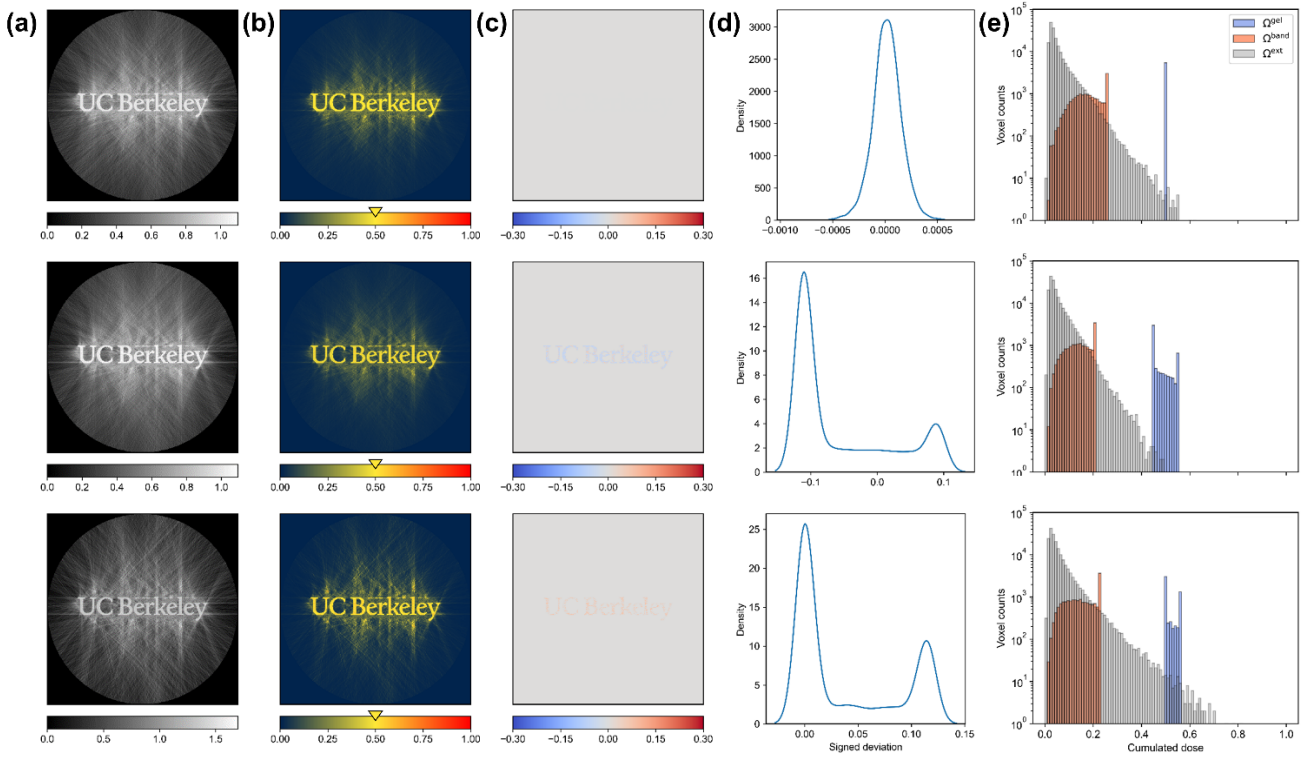

**Fig. 5.** Three LP formulations applied to a binary target (UC Berkeley logo) under different design objectives. Each row corresponds to a different formulation: **(first row)** general LP, **(second row)** target-bounded formulation (Case 1), and **(third row)** dose-shaping formulation (Case 2). Columns **(a–e)** present complementary views of the optimized solution: **(a)** optimized normalized dose $\tilde{\mathbf{f}}^*$, **(b)** resulting material response $\mathbf{m}^*$, **(c)** spatial map of signed response deviation relative to the target, **(d)** distribution of signed response deviation, and **(e)** histogram of accumulated dose for gelation $\Omega^{\text{gel}}$, band $\Omega^{\text{band}}$, and external $\Omega^{\text{ext}}$ regions. The comparison highlights the trade-off between in-part conformity and out-of-band dose suppression across the three formulations.

In contrast, Case 1 most aggressively suppressed dose spillage in the band region. This tendency is evident in Fig. 5(e), where the band distribution is shifted to lower values, and is quantitatively reflected by $\text{PSR}_m = 2.17$ and $\text{PSR}_f = 1.31$, together with reduced band maxima of $m^*_{\text{band,max}} = 0.21$ and $f^*_{\text{band,max}} = 0.37$. The improvement in process separation, however, came at the expense of target conformity. The uniformity metrics increased to $\text{DTVR}_m = 1.22$ and $\text{DTVR}_f = 1.09$, while the realized ratios in the gelation region spanned wider intervals of $0.899 \leq \frac{\mathbf{m}^*}{\mathbf{m}_T} \leq 1.101$ and $0.960 \leq \frac{\mathbf{f}^*}{\mathbf{f}_T} \leq 1.040$. As shown in Fig. 5(c,d), the deviation distribution is therefore substantially broader and extends to both positive and negative sides. This behavior is consistent with the role of Case 1 described in Sec. 2.7.1: the LP solver is not asked to reproduce the target center itself, but rather to remain within the admissible bounds while minimizing the worst-case dose spillage in band region $\Omega^{\text{band}}$.

Case 2 exhibited an intermediate behavior between these two extremes. Its target conformity was weaker than that of the general formulation but better than that of Case 1, with $\text{DTVR}_m = 1.13$ and $\text{DTVR}_f = 1.05$. At the same time, it preserved strong gel-band separation, with $\text{PSR}_m = 2.17$ and $\text{PSR}_f = 1.32$, which are nearly identical to the values obtained in Case 1. The realized ratios in the gelation region were $0.998 \leq \frac{\mathbf{m}^*}{\mathbf{m}_T} \leq 1.131$ and $0.999 \leq \frac{\mathbf{f}^*}{\mathbf{f}_T} \leq 1.053$, indicating that the solution remained tightly anchored to the lower bound while allowing a moderate upper-side overshoot. This one-sided pattern is consistent with the hard inhibition structure of Case 2: because the normalized solution is bounded below by the prescribed target, the LP tends to generate a systematic positive bias prior to post-scaling, as discussed in Sec. 2.7.2. After post-scaling, this bias is substantially reduced, which explains the narrower deviation profile in Fig. 5(d) relative to Case 1 while still maintaining the hard inhibition requirement in the band region $\Omega^{band}$.

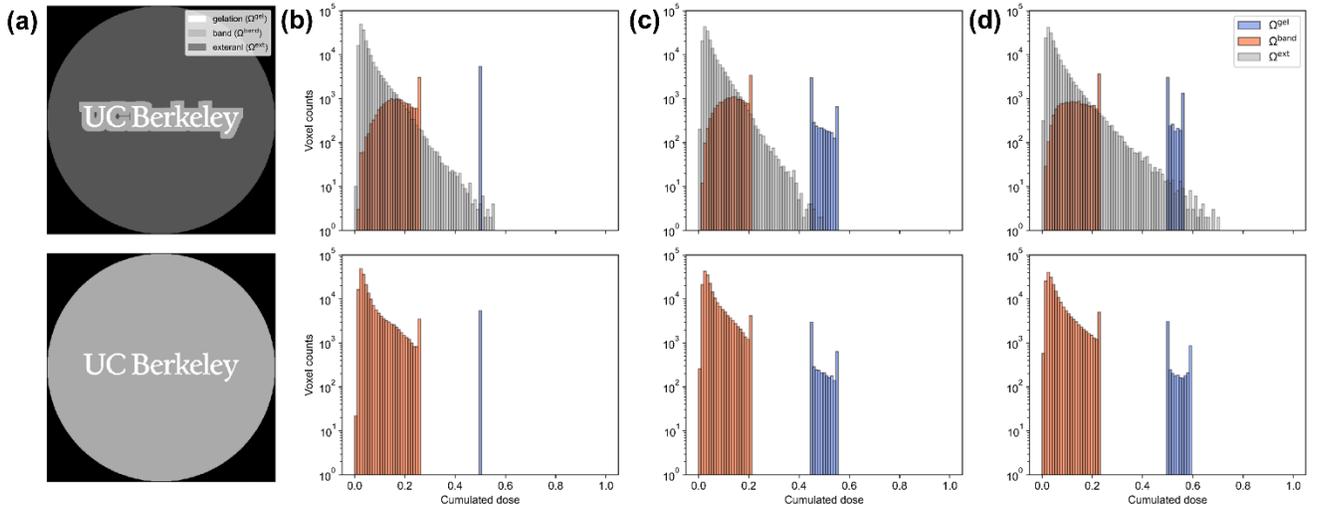

**Fig. 6.** Three LP formulations applied to a binary target (UC Berkeley logo) under different band-region definitions. **(top row)** the narrow surface band configuration with band width=10. **(bottom row)** the band free configuration. Within each row, columns **(b–d)** compare the three formulations: **(b)** general LP, **(c)** target-bounded formulation (Case 1), and **(d)** dose-shaping formulation (Case 2). Column **(a)** shows the target and region partition, while columns (b–d) present histograms of accumulated dose for the gelation $\Omega^{gel}$, band $\Omega^{band}$, and external $\Omega^{ext}$ regions. The comparison highlights how the band definition influences the separation between regions and alters the trade-off between in-part conformity and out-of-band dose suppression across the three formulations.

Overall, this binary benchmark shows that the three LP formulations should be interpreted as distinct operating modes rather than as a simple ranking of performance. The general formulation is most suitable when target conformity is the primary objective, Case 1 is most suitable when the widest process separation and strongest band suppression are required under strict admissible bounds, and Case 2 provides a compromise solution that preserves strong separation while recovering a significant portion of the target fidelity. In this comparison, PSR and DTVR are more informative than DSR for cross-case interpretation because the reference threshold differs among the formulations. In addition, Fig. 5(e) shows that non-negligible dose also appears in the external region $\Omega^{ext}$. This external-region spillage is a consequence of the present region-of-interest (RoI) definition, in which the LP explicitly penalizes dose only in the band region $\Omega^{band}$ surrounding the gelation part. As a result, the solver can create sacrificial spillage outside this enforced band while still improving the objective within the prescribed RoI. The effect of the band definition and the resulting sacrificial spillage are examined in the following section.

### 3.2. Effect of band-region definition on optimization behavior

To quantify the impact of band-region definition, we compare a narrow surface band (width = 10) with a band-free configuration in which the entire non-target domain is treated as $\Omega^{band}$. This comparison is important because conventional TVAM optimization is typically formulated without an explicit band region, whereas the present framework introduces $\Omega^{band}$ as an additional design variable for controlling out-of-bounds exposure. As shown in Fig. 6(b–d), removing the band causes only marginal changes in process separation: $PSR_m$ changes from 1.95 to 1.94 for the general formulation, from 2.17 to 2.15 for Case 1, and from 2.17 to 2.16 for Case 2. Thus, for the present Berkeley logo benchmark, a band-free formulation can still achieve nearly the same gel–void separation.

The main role of introducing a narrow band is therefore not a large gain in PSR, but a more controlled use of the limited feasible margin for sacrificial spillage. For the general formulation and Case 1, the width-10 case does not produce substantially larger external-region spillage than the band-free case, indicating that this geometry offers only a limited sacrificial margin outside the immediate boundary. In contrast, Case 2 shows a clearer benefit from the band definition. While its separation remains nearly unchanged, target conformity degrades when the band is removed: $DTVR_m$ increases from 1.13 to 1.18, corresponding to an approximately 4.2% loss in in-part uniformity. This indicates that, under the hard inhibition structure of Case 2, a tight boundary band helps preserve target conformity by restricting where residual spillage may occur.

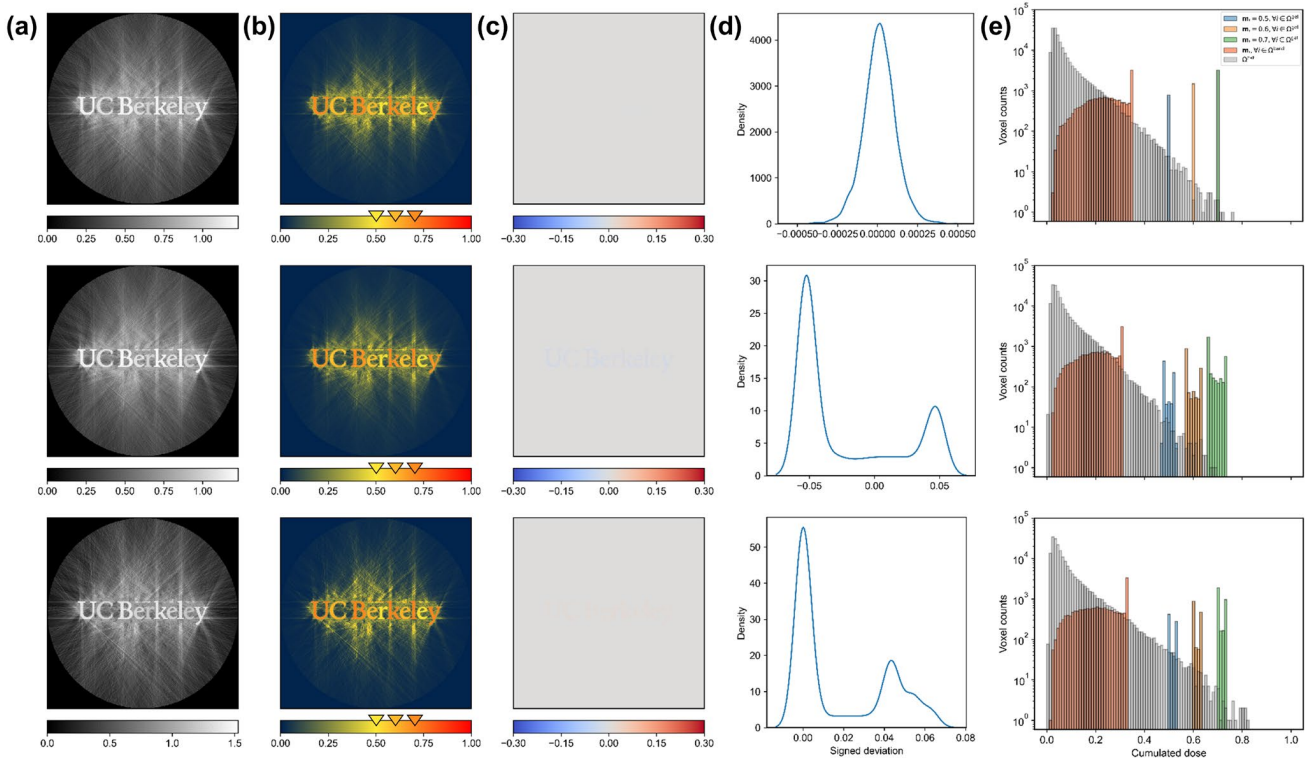

**Fig. 7.** Three LP formulations applied to a grayscale target (UC Berkeley logo) under different design objectives. Each row corresponds to a different formulation: **(first row)** general LP, **(second row)** target-bounded formulation (Case 1), and **(third row)** dose-shaping formulation (Case 2). Columns **(a–e)** present complementary views of the optimized solution: **(a)** optimized normalized dose $\tilde{\mathbf{f}}^*$, **(b)** resulting material response $\mathbf{m}^*$, **(c)** spatial map of signed response deviation relative to the target, **(d)** distribution of signed response deviation, and **(e)** histogram of accumulated dose for gelation $\Omega^{gel}$, band $\Omega^{band}$, and external $\Omega^{ext}$ regions with the prescribed grayscale target levels indicated for reference. The comparison underscores how the three formulations balance grayscale target conformity and out-of-band dose suppression.

Taken together, these results suggest that the benefit of an explicit band region is geometry-dependent. For simple targets such as the Berkeley logo, the optimizer can already achieve near-identical PSR without a dedicated band, implying that the available sacrificial margin is intrinsically small. Nevertheless, the Case 2 result demonstrates that even in such a setting, introducing a narrow band can improve target conformity without materially sacrificing process separation. This highlights the novelty of introducing $\Omega^{band}$ not merely as a spillage-control region, but as a design variable that shapes how the remaining optimization freedom is allocated.

From a practical standpoint, the band width should not be selected solely from an optimization perspective. In the present Berkeley logo benchmark, removing the band causes only marginal changes in PSR, which implies that the local gel–void separation can already be achieved without a dedicated boundary RoI. However, this does not mean that a band-free design is always physically safe. In real fabrication systems, residual sacrificial spillage outside the intended structure may induce unintended diffraction, scattering, or secondary curing, depending on the optical setup, resin sensitivity, and transport phenomena such as diffusion. For this reason, the minimum safe band width should be chosen in a physics-informed manner so that the imposed $\Omega^{band}$ is wide enough to capture the effective zone in which out-of-bounds exposure can still influence the final print. More broadly, the present results suggest that the optimal band definition is geometry- and process-dependent: when the available sacrificial margin is intrinsically small, as in the present target, a narrow band may already be sufficient, whereas more complex geometries or stronger physical spreading may require a wider and more conservative boundary region.

These observations also motivate a two-stage design strategy of multi-objective optimization. In the present results, Case 1 and Case 2 achieve nearly identical PSR values under both band definitions, yet Case 2 shows a clearer sensitivity in DTVR, particularly when the narrow band is removed. This indicates that the process-feasible separation bound can first be established independently of strict target centering, and only afterwards should the remaining optimization freedom be used to recover conformity. Based on this interpretation, one may first solve the target-bounded formulation (Case 1) to identify a physically feasible critical process bound under strict admissible limits, and then transfer that bound to the dose-shaping formulation (Case 2) to improve target conformity while preserving essentially the same separation level. In this sense, the first stage determines what process condition is safely achievable, whereas the second stage determines how well the target can be reproduced under that fixed condition. Such a sequential strategy is attractive because it is directly motivated by the observed decoupling between PSR and DTVR: the former remained nearly unchanged, while the latter still benefited from a tighter band definition in Case 2. Therefore, rather than relying on computationally expensive parameter sweeps, the proposed workflow suggests a practical route for process-aware LP design in which feasible process bounds are first identified and then exploited to recover fidelity in a controlled second stage.

### 3.3. Grayscale target with spatially varying prescription

Fig. 7 presents the three LP formulations applied to the segmented grayscale Berkeley logo target, where each character is assigned a distinct target response level. In this experiment, the band width was fixed at 10. For Case 1, the admissible response bounds were defined voxel-wise as $\mathbf{m}_L = (1-\epsilon_L)\mathbf{m}_T$ and $\mathbf{m}_U = (1+\epsilon_U)\mathbf{m}_T$ with $\epsilon_L = \epsilon_U = 0.05$, whereas Case 2 employed a hard inhibition threshold of $\mathbf{m}_{\text{crit}} = 0.33$ in the same manner in Sec. 3.1.

Among the three formulations, the general LP produced the highest grayscale target conformity. The response-domain uniformity metric reached $\text{DTVR}_m = 1.001$, and the realized response ratio in the gelation region remained tightly concentrated within $1.000 \leq \frac{\mathbf{m}^*}{\mathbf{m}_T} \leq 1.001$. This nearly ideal target-centered behavior is also evident in Fig. 7(c,d), where the signed normalized response deviation is sharply concentrated around zero. However, this fidelity was accompanied by the weakest gel-band separation, with $\text{PSR}_m = 1.41$ and $m^*_{\text{band,max}} = 0.35$. Thus, as in the binary benchmark, the general formulation remains strongly fidelity-oriented, but under grayscale prescription it operates with a reduced separation margin.

Case 1 exhibited the strongest separation at the expense of target conformity. It achieved the highest response-domain separation, $\text{PSR}_m = 1.55$, together with the lowest band maximum, $m^*_{\text{band,max}} = 0.31$. In contrast, the in-part uniformity degraded to $\text{DTVR}_m = 1.11$, and the realized response ratio spread over a much wider range, $0.949 \leq \frac{\mathbf{m}^*}{\mathbf{m}_T} \leq 1.051$. As shown in Fig. 7(c,d), the deviation profile is therefore substantially broader and extends to both positive and negative sides. This behavior is consistent with the role of Case 1: rather than reproducing the target center itself, the solver exploits the full admissible interval in order to maximize process separation under strict material tolerances.

Case 2 again provided an intermediate operating point between these two extremes. Its response-domain conformity was weaker than that of the general formulation but better than that of Case 1, with $\text{DTVR}_m = 1.07$, while still preserving strong separation with $\text{PSR}_m = 1.51$. The realized response ratio was $0.999 \leq \frac{\mathbf{m}^*}{\mathbf{m}_T} \leq 1.068$, indicating that the lower side remained tightly anchored to the target while only a moderate upper-side overshoot was allowed. This one-sided bias is consistent with the hard inhibition structure of Case 2 and mirrors the behavior observed in the binary benchmark.

Several observations follow from this grayscale benchmark. First, the overall separation level is lower than in the binary case for all three formulations, indicating that the available contrast margin is reduced when multiple prescribed response levels must be reproduced simultaneously. Second, the response-domain ratios $\frac{\mathbf{m}^*}{\mathbf{m}_T}$ provide a particularly clear interpretation of the formulation-dependent bias: the general LP remains target-centered, Case 1 fills the admissible interval to maximize separation, and Case 2 preserves the lower bound while allowing limited upper-side deviation. Third, despite the increased complexity of the grayscale target, the same qualitative hierarchy observed in the binary benchmark remains valid: the general formulation is fidelity-oriented, Case 1 is separation-oriented, and Case 2 provides a compromise solution under a hard process bound. These results support that the proposed LP framework remains interpretable and effective not only for binary targets but also for spatially varying grayscale prescriptions.

### 3.4. Extension to 3D with diffusion (physical forward model)

Whereas the preceding 2D benchmarks focused on the intrinsic behavior of the LP formulations under the idealized condition $\mathbf{K} = \mathbf{I}$, this subsection examines the validity of the proposed framework under a physically more realistic forward model with object-space blur, $\mathbf{A}^T = \mathbf{K}\mathbf{P}^T$. Fig. 8(a) shows the 3D Thinker target, and Fig. 8(c) shows the prescribed Gaussian PSF kernel embedded in the object-space operator $\mathbf{K}$. The purpose of this experiment is not to compare multiple formulations, but rather to test whether the general LP can still produce a process-feasible solution when diffusion-like spreading is explicitly included in the forward model.

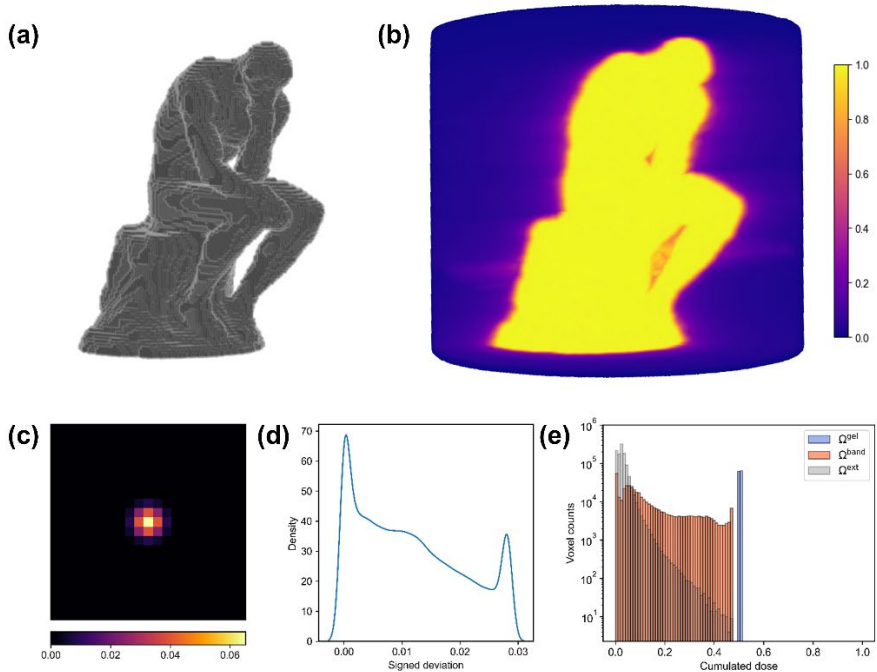

**Fig. 8.** Extension of the proposed framework to a physically realistic 3D setting with diffusion-like blur. **(a)** 3D Thinker target used for evaluation. **(b)** Normalized material response $\mathbf{m}_T / \max \mathbf{m}_T$ obtained by the general LP under the full forward model $\mathbf{A}^T = \mathbf{K}\mathbf{P}^T$ **(c)** Prescribed Gaussian PSF kernel used in the object-space operator $\mathbf{K}$. **(d)** Distribution of signed response deviation relative to the target within the gelation region. **(e)** Histogram of accumulated dose for the gelation $\Omega^{\text{gel}}$, band $\Omega^{\text{band}}$, and external $\Omega^{\text{ext}}$ regions. The result shows that the proposed framework maintains process-feasible separation even under 3D convolutional blur.

As shown in Fig. 8(b,d), the general LP still preserves a high level of target conformity despite the presence of 3D convolutional blur. In the material-response domain, the optimized solution yields $\text{DTVR}_m = 1.029$, while the realized response ratio in the gelation region remains within $1.000 \leq \frac{\mathbf{m}^*}{\mathbf{m}_T} \leq 1.029$. Thus, the minimum response is effectively pinned at the target level, with only a limited upper-side overshoot of approximately 3%. This indicates that the target-centered behavior observed for the general formulation in the 2D binary benchmark is largely maintained even under the full physical forward model. Consistently, the signed deviation distribution in Fig. 8(d) is concentrated in a narrow positive range, indicating that the solver continues to suppress lower-side under-response even in the presence of blur.

At the same time, the available process separation is markedly compressed relative to the 2D cases. In this experiment, $\text{PSR}_m = 1.06$, with extrema of $m^*_{\text{gel,min}} = 0.50$ and $m^*_{\text{band,max}} = 0.47$. Therefore, the gel-band separation remains process-feasible, in the sense that it stays above unity, but only with a very narrow margin. The histogram in Fig. 8(e) supports this interpretation: the gelation and band distributions lie much closer to one another than in the earlier 2D benchmarks, indicating that diffusion-like blur strongly compresses the feasible process window. In other words, the optimization freedom that could previously be exploited through band definition or sacrificial redistribution becomes substantially more limited once object-space spreading is introduced, and the trade-off between target conformity and process separation becomes governed more directly by the physical blur itself.

Interestingly, the histogram in Fig. 8(e) also shows that, even with ten voxels of band width, the external region $\Omega^{\text{ext}}$ remains relatively well separated, similar to the behavior observed in the band-free 2D optimization. This suggests that the 3D blur induced by the effective PSF smooths the gel-void interface itself, making it more difficult for the optimizer to accumulate locally elevated dose immediately outside the boundary. Rather than producing sharp boundary leakage, the object-space convolution distributes non-target exposure more broadly and smoothly. As a result, a narrow band already appears sufficient to capture the relevant worst-case boundary leakage in this setting. This observation implies that, when the $\mathbf{K}$ operator contains a sufficiently broad effective PSF, there may be little benefit in enlarging $\Omega^{\text{band}}$ at the expense of additional computational cost. In such blur-dominated physical forward models, the role of band definition may therefore be reduced relative to the 2D idealized setting.

## 4. Conclusion

This work presents a SiPO framework for TVAM. By formulating the problem in a normalized space, the proposed approach enables explicit control of target fidelity and out-of-bounds exposure for both binary and grayscale targets. The general formulation achieves near-ideal target conformity, while the target-bounded (Case 1) and dose-shaping (Case 2) formulations provide complementary operating regimes between process separation and fidelity. The introduction of a band region is shown to act as a design variable that governs the allocation of sacrificial spillage. Extension to a 3D forward model with diffusion demonstrates that physical blur significantly compresses the feasible process window, while the proposed framework remains capable of identifying process-feasible solutions. Finally, a two-stage optimization strategy is proposed to decouple feasibility identification from fidelity recovery, providing a practical alternative to parameter sweeps. Overall, this framework establishes a principled and scalable foundation for projection optimization in TVAM under physically realistic forward models.

# 6. Supplementary

## 6.1. Karush-Kuhn-Tucker (KKT) Conditions

Since the generalized scale-invariant formulation from Eq. (19) to Eq. (24) is a linear program, it is a convex optimization problem. Therefore, under standard feasibility assumptions, the KKT conditions are both necessary and sufficient for identifying the global optimum. Deriving these conditions not only provides the mathematical foundation for primal-dual optimization algorithms such as PDHG but also offers profound physical insights into how the solver manages dose distribution.

### 6.1.1. Lagrangian construction

Let $\mathbf{y}$, $\tilde{u}$, and $\tilde{v}$ be the primal variables. We introduce Lagrange multipliers (dual variables) for each constraint.

| Constraints | Lagrange multiplier | Space | Constraints description |
|---|---|---|---|
| $[\mathbf{A}^T\mathbf{y}]_i - \tilde{u} \leq 0$ | $\boldsymbol{\lambda}_i^{(1)} \geq 0$ | $\forall i \in \Omega^{\text{band}}$ | Auxiliary variable for upper bound in band region |
| $[\mathbf{A}^T\mathbf{y}]_i - \tilde{v}\tilde{\mathbf{f}}_{T,i} \leq 0$ | $\boldsymbol{\lambda}_i^{(2)} \geq 0$ | $\forall i \in \Omega^{\text{gel}}$ | Auxiliary variable for upper bound in gelation region |
| $\tilde{\mathbf{f}}_{T,i} - [\mathbf{A}^T\mathbf{y}]_i \leq 0$ | $\boldsymbol{\lambda}_i^{(3)} \geq 0$ | $\forall i \in \Omega^{\text{gel}}$ | Lower bound in gelation region |
| $-\mathbf{y}_j \leq 0$ | $\boldsymbol{\lambda}_j^{(4)} \geq 0$ | $\forall j \in \Pi^{\text{act}}$ | Lower bound for Active region (i.e., non-negativity) |
| $\mathbf{y}_j = 0$ | $\boldsymbol{\mu}_j^{(5)} \subset \mathbb{R}$ | $\forall j \in \Pi^{\text{mask}}$ | Masking for projection region |
| Primal feasibility* | Dual feasibility** | | |

The generalized Lagrangian function $\mathcal{L}(\mathbf{y}, \tilde{u}, \tilde{v}, \boldsymbol{\lambda}, \boldsymbol{\mu})$ is constructed as follows:

$$\mathcal{L}(\mathbf{y}, \tilde{u}, \tilde{v}, \boldsymbol{\lambda}, \boldsymbol{\mu}) = w_1 \tilde{u} + w_2 \tilde{v} + \sum_{i \in \Omega^{\text{band}}} \boldsymbol{\lambda}_i^{(1)} ([\mathbf{A}^T\mathbf{y}]_i - \tilde{u})$$

$$+ \sum_{i \in \Omega^{\text{gel}}} \boldsymbol{\lambda}_i^{(2)} ([\mathbf{A}^T\mathbf{y}]_i - \tilde{v}\tilde{\mathbf{f}}_{T,i})$$

$$+ \sum_{i \in \Omega^{\text{gel}}} \boldsymbol{\lambda}_i^{(3)} (\tilde{\mathbf{f}}_{T,i} - [\mathbf{A}^T\mathbf{y}]_i)$$

$$+ \sum_{j \in \Pi^{\text{act}}} \boldsymbol{\lambda}_j^{(4)} (-\mathbf{y}_j)$$

$$+ \sum_{j \in \Pi^{\text{mask}}} \boldsymbol{\mu}_j^{(5)} \mathbf{y}_j \quad (47)$$

### 6.1.2. Stationarity conditions

Setting the partial derivatives of $\mathcal{L}$ with respect to the primal variables to zero yields the stationarity conditions. For the scalar variables $\tilde{u}$ and $\tilde{v}$:

$$\frac{\partial \mathcal{L}}{\partial \tilde{u}} = w_1 + \sum_{i \in \Omega^{\text{band}}} \boldsymbol{\lambda}_i^{(1)}(-\mathbf{1}) = 0 \Rightarrow \sum_{i \in \Omega^{\text{band}}} \boldsymbol{\lambda}_i^{(1)} = w_1 \quad (48)$$

$$\frac{\partial \mathcal{L}}{\partial \tilde{v}} = w_2 + \sum_{i \in \Omega^{\text{gel}}} \boldsymbol{\lambda}_i^{(2)}(-\tilde{\mathbf{f}}_{T,i}) = 0 \Rightarrow \sum_{i \in \Omega^{\text{gel}}} \boldsymbol{\lambda}_i^{(2)} \tilde{\mathbf{f}}_{T,i} = w_2 \quad (49)$$

$$\frac{\partial \mathcal{L}}{\partial \mathbf{y}_j} = \sum_{i \in \Omega^{\text{band}}} \boldsymbol{\lambda}_i^{(1)} \mathbf{A}_{ji} + \sum_{i \in \Omega^{\text{gel}}} \boldsymbol{\lambda}_i^{(2)} (\mathbf{A}_{ji}) + \sum_{i \in \Omega^{\text{gel}}} \boldsymbol{\lambda}_i^{(3)} (-\mathbf{A}_{ji}) + \sum_{j \in \Pi^{\text{act}}} \boldsymbol{\lambda}_j^{(4)}(-\mathbf{1}) + \sum_{j \in \Pi^{\text{mask}}} \boldsymbol{\mu}_j^{(5)}(1)$$
$$= 0 \quad (50)$$

For the projection vector $\mathbf{y}$, the partial derivative with respect to each component $\mathbf{y}_j$ as given in Eq. (50) can be written in matrix form:

$$\nabla_{\mathbf{y}}\mathcal{L} = \mathbf{A}\left(\boldsymbol{\lambda}^{(1)}_{\Omega^{\text{band}}} + \boldsymbol{\lambda}^{(2)}_{\Omega^{\text{gel}}} - \boldsymbol{\lambda}^{(3)}_{\Omega^{\text{gel}}}\right) = \boldsymbol{\lambda}^{(4)}_{\Pi^{\text{act}}} - \boldsymbol{\mu}^{(5)}_{\Pi^{\text{mask}}} \tag{51}$$

### 6.1.3. Primal* and dual** feasibility

All original constraints must hold to satisfy primal feasibility. In terms of dual feasibility, Lagrange multipliers for inequality constraints are strictly non-negative $\boldsymbol{\lambda}^{(1)}_i, \boldsymbol{\lambda}^{(2)}_i, \boldsymbol{\lambda}^{(3)}_i, \boldsymbol{\lambda}^{(4)}_i \geq 0$, and the multiplier for the equality constraint $\boldsymbol{\mu}^{(5)}_j$ can be any real value.

### 6.1.4. Complementary slackness

Each constraint must be orthogonal to its multiplier. Note that equality constraint $\boldsymbol{\mu}^{(5)}$ always satisfies since its primal constraints must hold:

$$\begin{aligned}
\boldsymbol{\lambda}^{(1)}_i([\mathbf{A}^{\text{T}}\mathbf{y}]_i - \tilde{u}) &= 0 & \forall i \in \Omega^{\text{band}} \\
\boldsymbol{\lambda}^{(2)}_i([\mathbf{A}^{\text{T}}\mathbf{y}]_i - \tilde{v}\tilde{\mathbf{f}}_{T,i}) &= 0 & \forall i \in \Omega^{\text{gel}} \\
\boldsymbol{\lambda}^{(3)}_i(\tilde{\mathbf{f}}_{T,i} - [\mathbf{A}^{\text{T}}\mathbf{y}]_i) &= 0 & \forall i \in \Omega^{\text{gel}} \\
\boldsymbol{\lambda}^{(4)}_j \mathbf{y}_j &= 0 & \forall j \in \Pi^{\text{act}}
\end{aligned}$$